\documentclass[longauth] {aa}

\usepackage{txfonts}

\usepackage{graphicx}
\usepackage[numbers]{natbib}
\input epsf

\begin{document}

\title{The field high-amplitude SX~Phe variable BL~Cam: results from a
multisite photometric campaign
\thanks{Table\,2 is only available in
electronic form at the CDS via anonymous ftp to cdsarc.u-strasbg.fr
(130.79.128.5) or via
http://cdsweb.u-strasbg.fr/cgi-bin/qcat?J/A+A/}}

\subtitle{II. Evidence of a binary - possibly triple - system}

\author{ S. Fauvaud \inst{1,2}
\and J.-P. Sareyan \inst{3} \and I. Ribas \inst{4} \and E.
Rodr\'{i}guez \inst{5} \and P. Lampens \inst{6} \and G. Klingenberg
\inst{7}
 \and J. A. Farrell \inst{8}
\and  F. Fumagalli \inst{2} \and  J. H. Simonetti \inst {9} \and M.
Wolf \inst {10} \and G. Santacana \inst {1} \and  A.-Y. Zhou
\inst{11} \and R. Michel  \inst{12} \and L. Fox-Machado \inst{12}
\and M. Alvarez  \inst{12} \and A. Nava-Vega \inst{13} \and M. J.
L\'opez-Gonz\'alez \inst{5} \and V. M. Casanova \inst{5} \and F. J.
Aceituno \inst{5} \and I. Scheggia  \inst{14} \and J.-J. Rives
\inst{15} \and  E. G. Hintz \inst{16} \and  P. Van Cauteren
\inst{17} \and M. Helvaci \inst{18} \and C. Yesilyaprak \inst{19}
\and K. A. Graham \inst{20} \and L. Kr\'al \inst{21} \and  R.
Koci\'an \inst{21} \and H. Ku\v{c}\'akov\'a \inst{21} \and M.
Fauvaud \inst{1} \and B. H. Granslo \inst{7} \and J. Michelet
\inst{22} \and  M. P. Nicholson \inst{20} \and J.-M. Vugnon
\inst{23} \and  L. Kotkov\'a \inst{24} \and K. Truparov\'a \inst{21}
\and C. Ulusoy \inst{25} \and  B. Yasarsoy \inst{25} \and A.
Avdibegovic \inst{10} \and M. Bla\u{z}ek \inst{10} \and J. Kliner
\inst{10} \and P. Zasche \inst{10} \and  S. Barto\v{s}\'{i}kov\'a
\inst{21} \and M. Vil\'a\v{s}ek  \inst{21} \and  O. Trondal \inst{7}
\and F. Van Den Abbeel \inst{26} \and R. Behrend \inst{27} \and H.
W\"{u}cher \inst{1} }

\institute{Observatoire du Bois de Bardon, 16110 Taponnat, France,
e-mail: stephane.fauvaud@wanadoo.fr \and Groupe Europ\'een
d'Observation Stellaire (GEOS), 23 Parc de Levesville, 28300
Bailleau l'Ev\^{e}que, France \and  Observatoire de la C\^{o}te
d'Azur, BP 4229, 06304 Nice cedex 4, France, and Observatoire de
Paris-Meudon, LESIA, 92190, Meudon, France, e-mail:
jean-pierre.sareyan@obspm.fr  \and Institut de Ci\'encies de l'Espai
(CSIC-IEEC), Campus UAB, Facultat de Ci\'encies, Torre C5, parell,
2a pl., E-08193 Bellaterra, Spain \and  Instituto de Astrof\'{i}sica
de Andaluc\'{i}a, CSIC, PO Box 3004, 18080 Granada, Spain \and
Koninklijke Sterrenwacht van Belgi\"{e}, Ringlaan 3, 1180 Brussel,
Belgium \and  Variable Star Section, Norwegian Astronomical Society,
PO Box 1029 Blindern, 0315 Oslo, Norway \and Sulphur Flats
Observatory, 449 Sulphur Creek Road, Jemez Springs, NM 87025, USA
\and Martin Observatory, Physics Dep., Virginia Polytechnic
Institute \& State Univ., Blacksburg, VA 24061, USA \and
Astronomical Institute, Charles University Prague, V Holesovickach
2, 180 00 Praha 8, Czech Republic \and National Astronomical
Observatories, Chinese Academy of Sciences, Beijing 100012 China
        \and Observatorio Astron\'omico Nacional, Instituto de
Astronom\'{\i}a,  Universidad Nacional Aut\'onoma de M\'exico, A.P.
877, Ensenada, BC 22860, M\'exico \and  Facultad de Ciencias
Qu\'{i}micas e Ingenier\'{i}a, Universidad Aut\'onoma de Baja
California, Calzada Tecnolog\'{i}co No. 14418, Mesa de Otay, C.P.
22390, Tijuana, B.C. M\'exico \and Societ\'a Astronomica Le pleiadi,
Italy \and Observatoire des trois korrigans, 85600 Treize-Septiers,
France \and  Department of Physics and Astronomy, Brigham Young
University, N283 ESC, Provo, UT 84602, USA \and Beersel Hills
Observatory, B-Beersel, BelgiumÂ \and  Akdeniz University,
Department of Physics, 07058 Antalya, Turkey \and Ataturk
University, Department of Physics, 25340 Erzurum, Turkey \and
American Association of Variable Star Observers, 25 Birch St.,
Cambridge, MA 02138, USA
         \and  Project Eridanus, Observatory and Planetarium of Johann Palisa,
Tr. 17. listopadu 15, 708 33 Ostrava-Poruba, Czech Republic \and
Club d'Astronomie Lyon Amp\`ere, 37 rue Paul Cazeneuve, 69008 Lyon,
France \and Club Eclipse, 22 rue du Borr\'ego, Appt 5A4, 75020
Paris, France \and  Astronomical Institute, Academy of Sciences, 251
65 Ondrejov, Czech Republic \and  Ege University Observatory, 35100
Bornova, Izmir, Turkey \and Vesqueville Observatory, rue de Fayet 8,
B-6870 Vesqueville, Belgium \and Geneva Observatory, CH-1290
Sauverny, Switzerland  }

\date{Received  /  Accepted }

\authorrunning{S. Fauvaud et al.}
\titlerunning{Multiplicity in BL~Cam}

\abstract
{Short-period high-amplitude pulsating stars of Population I
($\delta$~Sct stars) and II (SX~Phe variables) exist in the lower
part of the classical (Cepheid) instability strip. Most of them have
very simple pulsational behaviours, only one or two radial modes
being excited. Nevertheless, BL~Cam is a unique object among them,
being an extreme metal-deficient field high-amplitude SX~Phe
variable with a large number of frequencies. Based on a frequency
analysis, a pulsational interpretation was previously given.}
{We attempt to interpret the long-term behaviour of the residuals
that were not taken into account in the previous Observed-Calculated
(O-C) short-term analyses.}
{An investigation of the O-C times has been carried out, using a
data set based on the previous published times of light maxima,
largely enriched by those obtained during an intensive multisite
photometric campaign of BL~Cam lasting several months.}
{In addition to a positive (161 $\pm$ 3) x 10$^{-9}$~yr$^{-1}$
secular relative increase in the main pulsation period of BL~Cam, we
detected in the O-C data short- (144.2 d) and long-term ($\sim$ 3400
d) variations, both incompatible with a scenario of stellar
evolution.}
{Interpreted as a light travel-time effect, the short-term O-C
variation is indicative of a massive stellar component (0.46 to 1
M$_{\sun}$) with a short period orbit (144.2 d), within a distance
of 0.7 AU from the primary. More observations are needed to confirm
the long-term O-C variations: if they were also to be caused by a
light travel-time effect, they could be interpreted in terms of a
third component, in this case probably a brown dwarf star ($\geq$
0.03 \ M$_{\sun}$), orbiting in $\sim$ 3400 d at a distance of 4.5
AU from the primary.}


\maketitle


\section{Introduction}

SX~Phe-type stars of the Galactic field are high-amplitude
Population II pulsators located in the lower part of the classical
(Cepheid) instability strip, close to the main sequence, among the
high-amplitude Population I $\delta$
 Scuti variables (e.g., Nemec \& Mateo 1990; Breger 2000; Rodr\'{i}guez 2003).
Only a few (14) field SX~Phe-type stars have been found to date.
However, more than one hundred SX~Phe variables have been discovered
among the blue straggler population of globular clusters, and in
nearby dwarf galaxies (e.g., Rodr\'{i}guez \& L\'opez-Gonz\'alez
2000; Jeon et al. 2003, 2004; Mazur et al. 2003).

Multiperiodicity with more than two independent modes is not a
common feature among the high-amplitude pulsators of the low
instability strip. However, BL~Cam (= GD 428, $\alpha_{2000}$ = 3 h
47 m 19 s, $\delta_{2000} = + 63\degr 22\arcmin  07\arcsec$,
$<$V$>$ $\sim$ 13.1 mag, $\Delta V=0.33$ mag) is one of these very
few exceptions. In addition to its main pulsation frequency $f_0 =
25.5769$ cycles per day (d$^{-1}$), the multiperiodicity of this
extreme metal-deficient ([Fe/H] = -2.4 dex; McNamara 1997) field
SX~Phe-type star has been claimed by different authors (cf. Fauvaud
et al. 2006  for a review, hereafter F06). Analysis of the main
period temporal changes of BL~Cam has also revealed a secular
increase in its main period and the possibility that this star is a
member of a detached binary system (F06).

Since BL~Cam is a very attractive target for observational
asteroseismology, we initiated the first multisite photometric
campaign for this object to investigate in detail its pulsational
behaviour. The results of a frequency analysis (Rodr\'{i}guez et al.
2007, hereafter R07) has largely allowed the reduction of the daily
alias and provided the true value of the first overtone frequency
(32.6~d$^{-1}$). R07 confirmed the existence of a very dense
pulsational frequency spectrum in this star, in addition to the
already known high-amplitude main periodicity.

In this paper, we present a new analysis of the main period
variation of BL~Cam by means of the Observed-Calculated (O-C)
diagram of the O-C times. From a general point of view, the O-C
method (see, e.g., Sterken 2005a) allows us to assess the
consistency between the measurement of an observable phenomenon and
its prediction, and the possible importance of the light travel-time
(LTT) effect in the O-C diagram. A historical example of this method
is the discovery made by Ola\"{u}s R\"{o}mer, in 1676, of the finite
velocity of light, obtained observing the eclipse phenomena of
Jupiter's Io satellite (D\'{e}barbat 1978; Sterken 2005b). The
variable LTT effect, caused by the different positions of the Earth
on its orbit, makes it necessary to convert Julian dates into
Heliocentric Julian dates. In variable star studies, the analysis of
the O-C diagram is a widely used tool.

The paper is organized as follows. The observations and data
reduction are presented in Sect. 2. A new ephemeris and a
description of the O-C diagram are given in Sect. 3. An
interpretation of the O-C values is proposed in Sect. 4, and Sect. 5
is devoted to a discussion. A conclusion and perspectives are given
in Sect. 6.

\section{Observations and data reduction}

Our multisite photometric campaign was performed between August 2005
and February 2009, from 24 professional and amateur observatories
located in Europe and North America. The participating sites from
August 2005 to March 2007, their instrumentation and the methods of
data reduction, including the selected comparison stars, were
described in R07.

From October 2007 to February 2009, new photometric observations of
BL~Cam were carried out mainly from Observatorio Astron\'omico
Nacional San Pedro Mart\'{i}r (Baja California, M\'exico), Calina
Observatory (Switzerland), and Observatoire du Pic du Midi (France).
Some other data were collected at stations located in Belgium, Czech
Republic, France, and USA. Table\,1 presents a summary of the
observations with the list of the observatories. All data were
obtained with CCD cameras using various filters, including some
observations without any filter.

\begin{table*}
\begin{center}
\caption[]{Journal of observations carried out from October 2007 to
February 2009.} \label{Table 1}
\begin{tabular}{l c c c c c c} \hline \hline
Observatory & Location & Telescope(m) & Filters & Nights & Hours \\
\hline SPM & San\ Pedro \ M\'artir (M\'exico)&  0.84  & \it{Vby}&
22&  97.8\tablefootmark{a} \\
Cal &Carona (Switzerland)         &  0.14/0.30 &      no&  14   &  81.1 \\
PiM & Pic\  du\  Midi (France)     &  0.60      & V&   6 &  34.9\\
O3k &  Treize-Septiers (France)     &  0.20      & no&   6&  19.5\tablefootmark{b} \\
Ond &  Ondrejov (Czech Republic)     &  0.65      &R&   6&  14.9 \\
Gil &  Gilman (USA)            &  0.30   &V&   2&   4.8 \\
Mar &Blacksburg (USA)     &  0.35      &V&   2&   4.4 \\
Gras&Mayhill (USA)             &  0.30 &     R&   1&   2.2 \\
Ves &Vesqueville (Belgium)    &  0.20  &V&   1&   1.9 \\
\hline
\end{tabular}
\end{center}
\tablefoot{Observatory code: Cal = Calina Observatory, Gil = Gilman
Observatory, Gras = Global Rent-a-Scope Observatory, Mar = Martin
Observatory, O3k = Observatoire des trois korrigans, Ond=Ondrejov
Observatory, PiM = Observatoire du Pic du Midi, SPM =
San Pedro Mart\'{i}r Observatory, Ves = Vesqueville Observatory.\\
\tablefoottext{a}{Str\"omgren $\textit{b}$ and $\textit{y}$ filters
were used during one night.} \tablefoottext{b}{V filter was using
during 2 hours}. }
\end{table*}

\setcounter{table}{2}

The times of the observed light maximum were determined by fitting
each peak of the light curves by a third or a fifth degree
polynomial, using the software package Peranso (Vanmunster 2008). We
estimate the standard deviation (1$\sigma$) in an individual maximum
to be between $\sim$0.0003 and 0.0006~d ($\sim$26 and 52 s).
Comparing the maximum times sets obtained by different observers
during simultaneous runs indicates, however, a scatter of between
$\sim$0.0001 and 0.002 d ($\sim$10 and 200 s), yielding a typical
1$\sigma$ uncertainty of $\sim$0.0008 d (i.e., $\sim$70 s).

A total of 930 new times of light maximum, listed in Table\,2, were
recorded  during our runs. Table\,2 also includes 9 maximum times
based on measurements from the AAVSO database, and 20 maxima that we
estimated from light curves of Martin (1996, 2001). Our new analysis
of the O-C residuals for the main period of BL~Cam is based on a
data set consisting of 1510 times of light maximum, constituted by
those used in F06 (with the 61 unpublished maxima from Delaney et
al. 2000), including the light maxima of Table\,2, and an additional
2 and 73 maxima published in Klingenberg et al. (2006) and Fu et al.
(2008a), respectively. Our complete data set spans from December
1976 to February 2009, i.e., more than 30 years.

Even though not explicitly mentioned, the light maximum times
published in previous papers are commonly assumed to be in
coordinated universal time (UTC). However, the fluctuations of the
Earth's rotation rate (e.g., Souchay 2007; McCarthy \& Seidelmann
2009) affect UTC, which runs with irregular jumps of a full second.
Over the past 30 years, the deviation of UTC from a fixed frame of
time reference has increased in steps of 1~s, making it roughly an
average of 1~s~yr$^{-1}$. The consequences become significant for
LTT effects when a time baseline longer than a few years is
considered, and when the amplitude of the effect is small (Ribas
2005). In addition, the secular decrease rate of the Earth's spin
induces a slow decrease in the variation in the pulsation period of
a star (Bastian 2000). To avoid these problems, we converted all UTC
timings into barycentric dynamical time (TDB), using the expressions
of Seidelmann \& Fukushima (1992) and corrections provided by
Berthier (2009 private communication). Thus, TDB-UTC varies from $+
47.184$~s in 1976 to $+66.184$~s in 2009.

Table\,2 gives both the calculated $HJD_{max}$ and $BJD_{max}$ times
of light maximum, expressed in Heliocentric Julian Days (HJD in the
UTC scale) and Barycentric Julian Days (BJD, i.e. HJD in the TDB
scale), respectively. However, our analysis of the O-C diagram was
performed using $BJD_{max}$.

\section{The O-C diagram }

A weight ($\frac{1}{\sigma^2}$) was assigned to each time of maximum
for the data set described in Sect. 2. When no uncertainty was
published, a median uncertainty $\sigma$ = 0.0005 d was adopted
(F06). Hereafter, $E$ is the cycle number elapsed since the initial
epoch of the ephemeris (i.e., JD$\sim$2443125.8).

We first checked and revised the linear ephemeris of Zhou et al.
(1999, Eq.~2). The best-fit least squares regression to the entire
weighted light maxima data set gives the new linear ephemeris

\begin{equation}
BJD_{max} = BJD\ 2443125.80068(4) + 0.0390978897(2)E
\end{equation}

~~~~
\newline
which has a standard deviation of $\sigma$  = 0.0025 d (the
uncertainty in the last digit is indicated in parentheses).

\begin{figure}
\begin{center}
\epsfxsize=8.0truecm \epsfbox{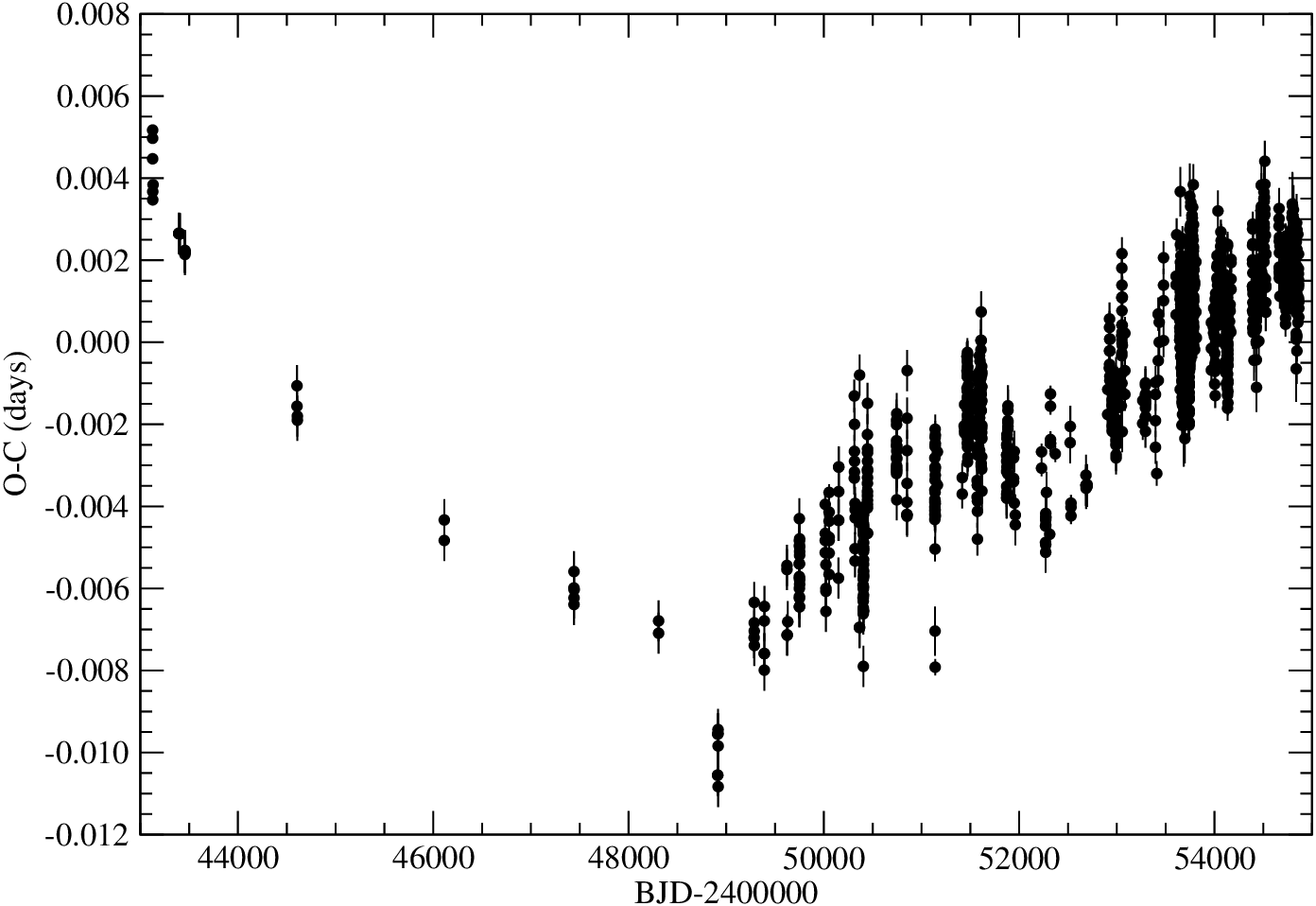} \caption[]{O-C diagram for
BL~Cam using the new linear ephemeris (Eq.~1).}
\end{center}
\end{figure}

Despite the revision of Zhou et al.'s linear ephemeris, a parabolic
trend remains in the O-C diagram (Fig.\,1), and the period and
initial epoch do not match the light maxima. The parabolic shape
indicates that the main period of BL~Cam slowly increased with time,
which can be represented by a second-order polynomial. The best-fit
relation of the data yields the new, more accurate, quadratic
ephemeris

\begin{eqnarray}
    BJD_{max} &  = & BJD \ 2443125.80466(6) +  0.0390977846(6) E \nonumber  \\
              &    &   + 3.367(15) \times 10^{-13} E^2
\end{eqnarray}

~~~
\newline
The standard deviation is $\sigma$~=~0.0013~d (Fig.\,2). The squared
term implies a secular increase rate $dP/(Pdt)$ =
(4.41~$\pm$~0.06)~x~10$^{-10}$~d$^{-1}$ =
(161~$\pm$~3)~x~10$^{-9}$~yr$^{-1}$ of the main pulsation period P
of BL~Cam with time $t$, i.e., a period increase of about 1~s per
1800~yr.

\begin{figure}
\begin{center}
\epsfxsize=8.0truecm \epsfbox{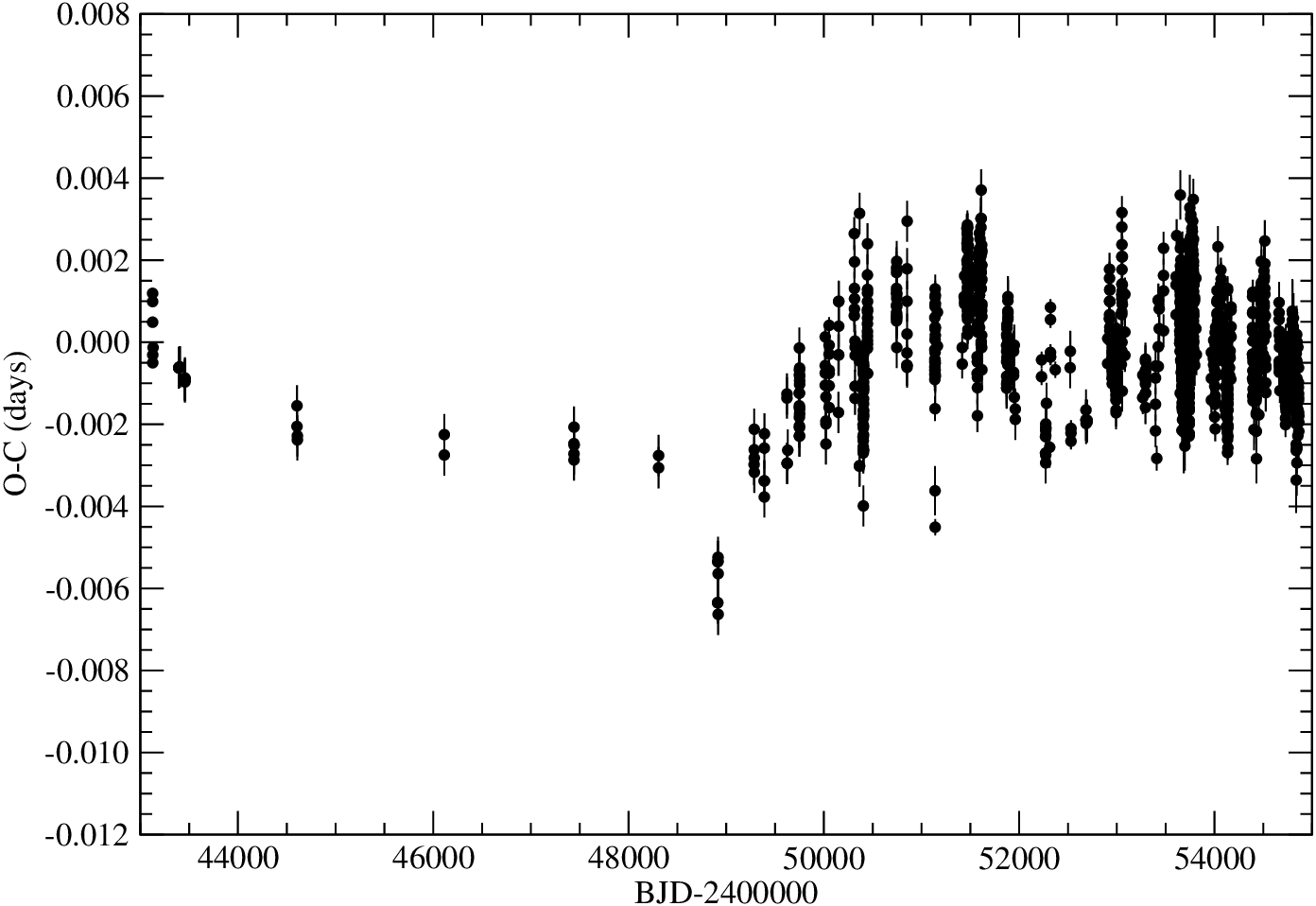} \caption[]{O-C diagram for
BL~Cam using the new quadratic ephemeris (Eq.~2).}
\end{center}
\end{figure}

If we remove the parabolic shape, significant residuals remain in
the O-C values (Fig.\,2), which have a long periodic modulation, of
a timescale of several years. These residuals have already been
reported and interpreted as a LTT effect (with a semi-amplitude of
148~$\pm$~12~s and a period of 10.5~$\pm$~0.2~yr, according to F06),
i.e., a cyclic light timing signal as received by a terrestrial
observer, due to the wobble of the main star's barycentre induced by
a companion.

If we focus on the period between August 2005 and February 2009,
these new intensive observational run data indicate that a
short-term ($\sim$0.4~yr) periodic O-C shift also appears. A careful
inspection of the O-C diagram shows that this behaviour was present
in the past. No single observation conflicts with this observed O-C
oscillation. Indeed, this short-term variation is:

\begin{itemize}
\item{obvious during our multisite campaign (Fig.\,3), i.e., between August 2005
and February 2009 (JD 2453606 to 2454863);}
\item{probable between September 2003 and March 2004 (JD 2452906 to 2453085);}
\item{very possible between August 1999 and March 2000 (JD 2451415 to
2451620), and from August 1996 to January 1997 (JD 2450310 to
2450450).}
\end{itemize}

\begin{figure}
\begin{center}
\epsfxsize=8.0truecm \epsfbox{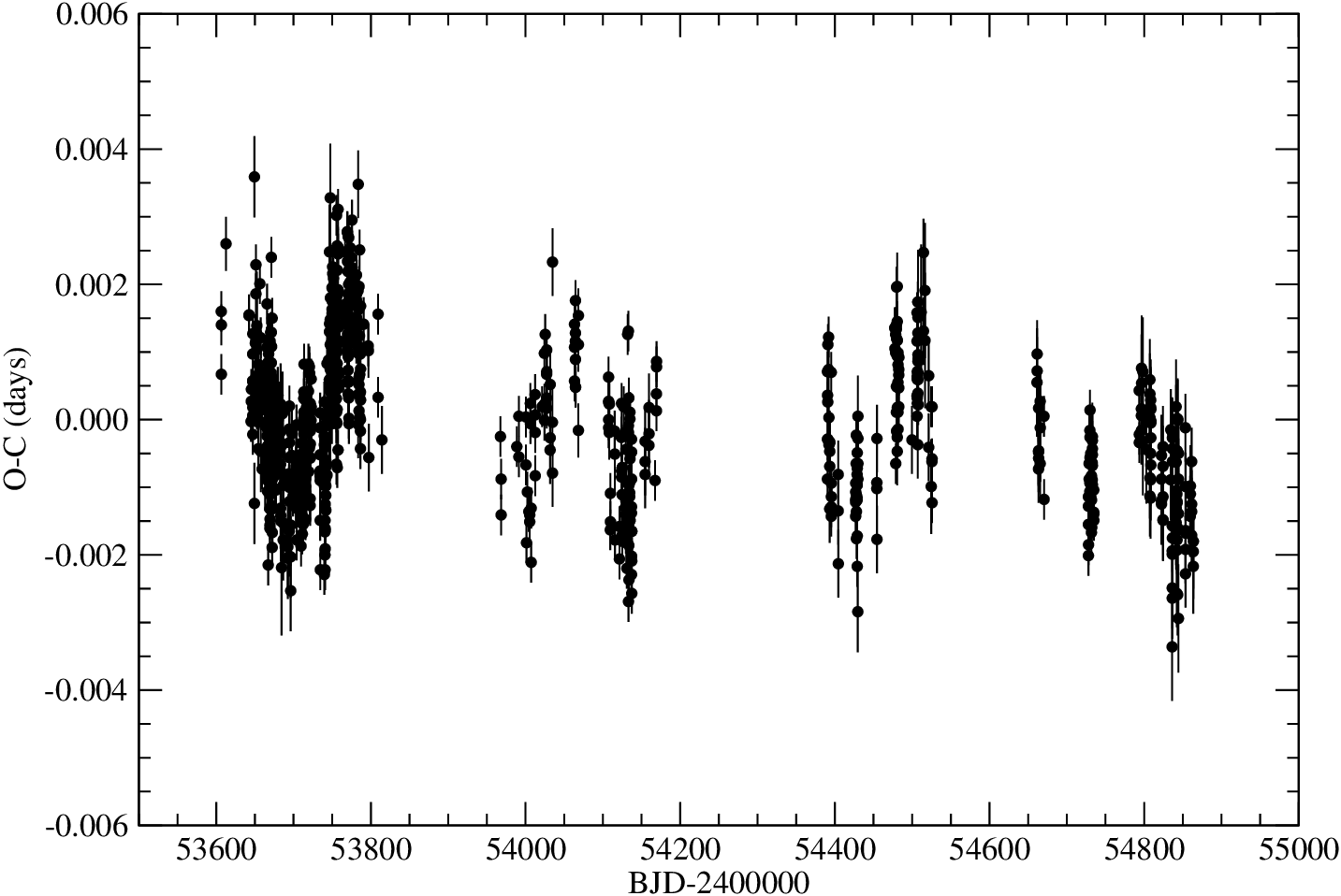} \caption[]{Detail of Fig.~2
(August 2005 to February 2009), showing a periodic short-term
variation.}
\end{center}
\end{figure}

These long- and short-term modulations -- and their large amplitude
residuals -- are compatible with neither stellar evolution on a
secular timescale nor the timescale of the pulsations ($\sim$1
hour). As period changes in pulsating variable stars are related to
the mean density variation of a star during its evolution (Eddington
1918), this basic interpretation is insufficient to explain the
observations in numerous cases (e.g, Handler 2000). For the SX~Phe
variables, the discrepancy between the theoretical period changes
and the observed rates shows that non-evolutionary or non-linear
processes must be invoked (e.g., Rodr\'{i}guez 2003 for a review).
On the one hand, several physical processes have been suggested to
explain the observed non-evolutionary period changes, for example
stellar companions, and non-linear effects in pulsation caused by
stellar rotation or resonant coupling frequencies (e.g., Breger \&
Pamyatnykh 1998; Szeidl 2005; Templeton 2005). On the other hand,
according to a study of Percy et al. (2007) of three SX~Phe-type
stars, the contribution of random cycle-to-cycle period fluctuations
seems insufficient to explain the non-parabolic shape of the O-C
diagram.

While the parabolic main trend of the O-C diagram of BL Cam is
likely a consequence of a period increase due to secular evolution,
the time constants ($\sim$10 and 0.4~yr) involved in the long- and
short-term oscillations point at non-evolutionary processes.
Although adjustments in the internal structure of the star cannot be
completely ruled out to generate abrupt period changes (see F06),
our simple interpretation that we propose here is that two bodies
orbit BL~Cam, producing LTT effects in its O-C diagram.

\section{BL~Cam as a binary system, and the possible existence of a
third body}

A Fourier analysis of the O-C values, obtained from Eq. 2, was
carried out with the Period04 software developed by Lenz \& Breger
(2005). The long-term modulation was investigated, as well as the
short one in the JD ranges described in Sect. 3, with a
0.01~d$^{-1}$ frequency step. Two periods were found, $\sim$7~yr and
$\sim$144 d, with semi-amplitudes $\sim$52 s and $\sim$105 s,
respectively. However, the oldest data (JD $<$ 2449000) are sparse
or might have large uncertainties. Therefore, our Fourier analysis
focused on the data obtained after JDÂ \  2449000, which infers
periods of $\sim$8.6~yr and 144.18 d, with respective
semi-amplitudes $\sim$82 s and 102.6 s.

A three-body model fit of the O-C diagram was then performed on the
highest quality O-C data (JD $ > $ 2449000). Historically, such a
triple system was first suggested by Chandler (1892) for Algol,
precisely based on the assumption of  LTT effects.

The basic equations describing the LTT effect of a binary system as
a function of the orbital motion were first proposed by Irwin
(1952). The formalism used in this study was described in Ribas et
al. (2002); these authors developed a model that takes into account
all the binary parameters such as semi-amplitude, eccentricity,
argument of periastron, orbital period, and time of passage at
periastron.

Since our LTT effect computing program can only deal with one orbit
at a time, we used an iterative process where we first introduced
the  short orbital period. The fitting of the data started with the
period and semi-amplitude obtained from the Fourier analysis. The
short period orbit was first considered to be circular, and when all
the other parameters are settled, the computation of the
eccentricity was carried out.

In the case of the long period orbit, the residuals in the short
period LTT effect were first computed, and we then started a new
computation, based on the values derived from the previous Fourier
analysis. We then searched for solutions of fixed eccentricity by
scanning different values (leaving all other parameters free). An
independent attempt to fit the data, leaving all parameters free,
was unsuccessful.

Table\,3 presents the final results of the best fit obtained for the
LTT effect caused by the short orbit, when considering JD $>$
2449000 data. The standard deviation of the fit is 78.8 s. The
144.19 $\pm$ 0.02 d periodicity is found for BL~Cam~B, assuming  a
mass of 0.99 M$_{\sun}$ for BL~Cam~A (McNamara 1997). Fig.\,4 shows
the phase-folded fit corresponding to this short period.

If it were caused a third object, a LTT effect analysis would infer
a long orbit consistent with a rather high eccentricity ($>$ 0.6), a
$\sim$9.3~yr period, and a semi-amplitude around 60 s (this last
value being especially sensitive to the eccentricity). An
eccentricity of 0.7 yields a 3390 $\pm$ 26 d period (i.e. 9.28
$\pm$~0.07~yr, with a 65.6~$\pm$~1.2~s semi-amplitude). Although a
long oscillation is clearly visible in the O-C diagram and is
unlikely to be spurious, the data cover only about two cycles,
meaning that this LTT model (Fig.\,5) has a weaker significance
compared to the short 144 d orbit already found.

\begin{figure}
\begin{center}
\epsfxsize=7.0truecm \epsfbox{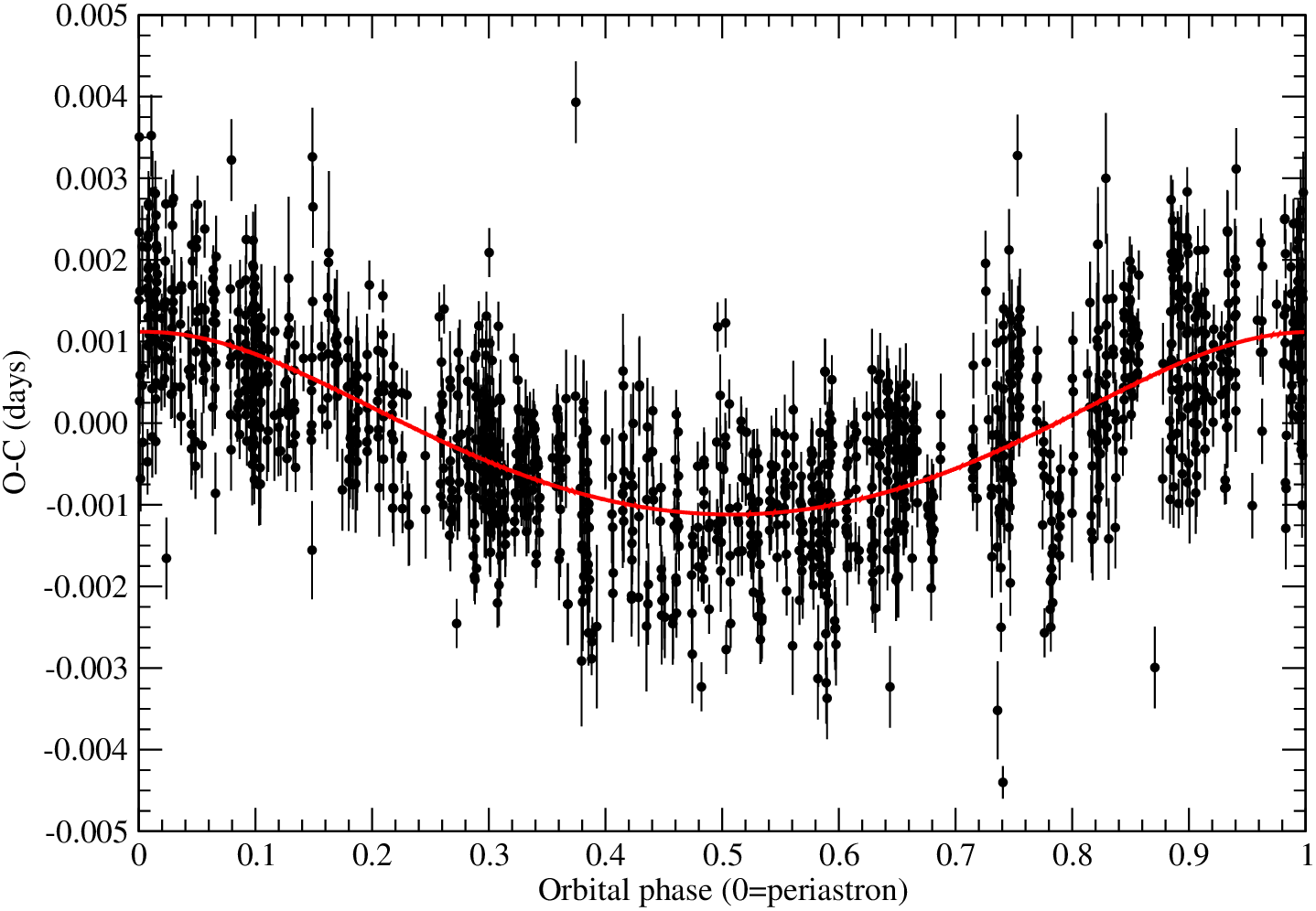} \caption[]{O-C
diagram for BL~Cam phased against the short orbital period
(144.19$\pm$0.02~d), according to the best-fit LTT model
(Table\,3).}
\end{center}
\end{figure}

\begin{figure}
\begin{center}
\epsfxsize=7.0truecm \epsfbox{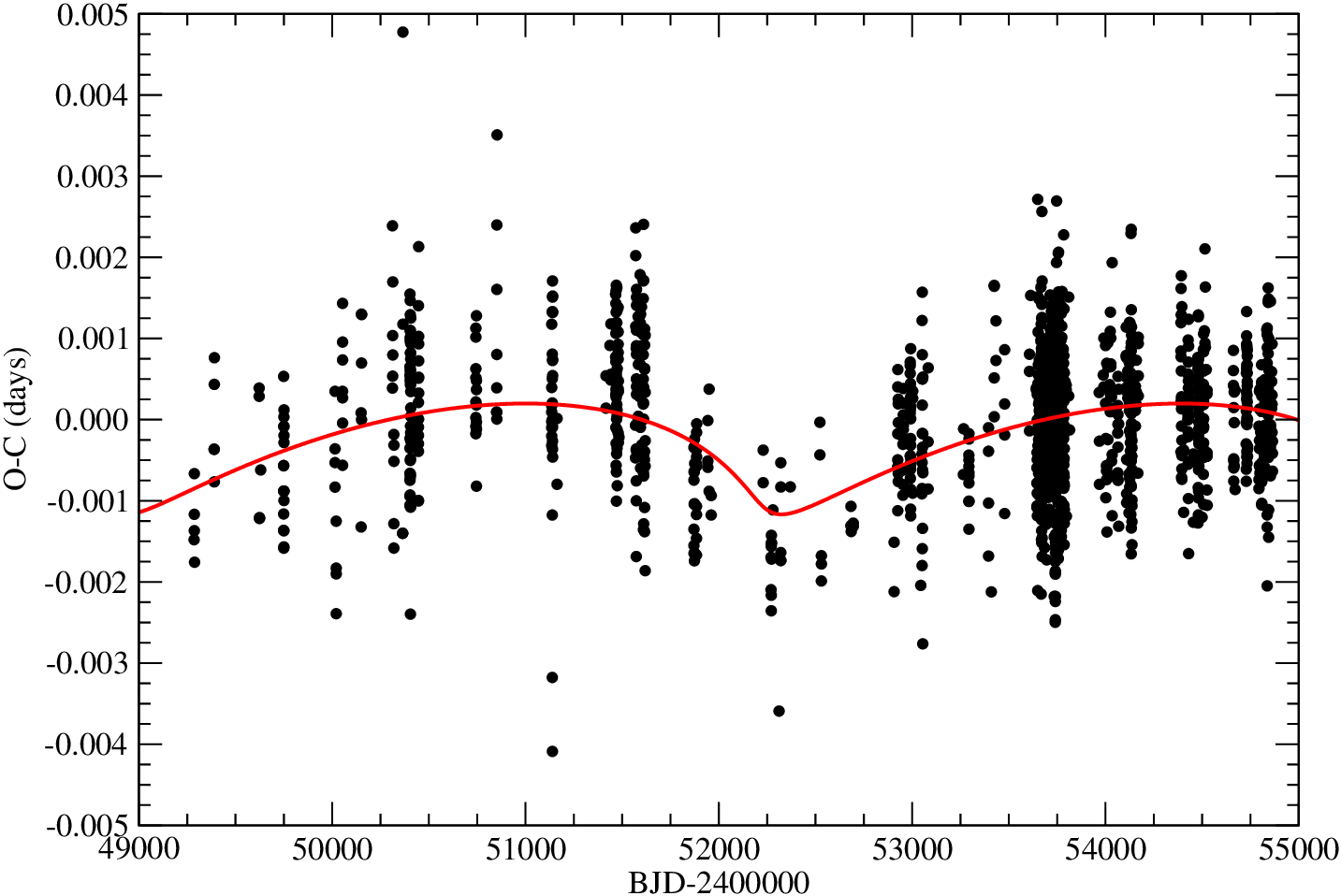} \caption[]{O-C
diagram for BL~Cam phased against the long orbital period
(3390$\pm$26~d) fitting, according to the best-fit LTT model.}
\end{center}
\end{figure}

These large O-C variations could also be accounted for by successive
abrupt period changes (i.e., straight segments in the O-C diagram,
instead of a stable periodic variation). As mentioned before, our
actual phase coverage is insufficient to establish an unambiguous
interpretation. In the case of a third (BL~Cam~C) object, its
derived orbital properties strongly depend on the long-term
correction of the O-C trends (either linear or quadratic). Thus this
possibility has to be confirmed by future continuous monitoring of
the star. In the following steps of this study, we decided not to
rule out the hypothesis of a second companion, even if only
suspected.

Since we do not observe any eclipse in our photometric observations
of BL~Cam, the inclination of the system cannot be precisely
determined. However, the Kepler's third law can be used to estimate
the minimum masses and minimum semi-major axes of the two companions
BL~Cam~B and BL~Cam~C, BL~Cam~A being the primary and SX~Phe
variable star of the system (e.g., Hilditch 2001). Indeed, assuming
a mass of 0.99~M$_{\sun}$ for BL~Cam~A, BL~Cam~B should have a mass
m$_B$~$\geq$~0.46~M$_{\sun}$ and its orbit a semi-major axis
a$_B$~$\geq$~0.6~AU; BL~Cam~C should be less massive than BL~Cam~B,
with a mass m$_{C}$~$\geq$~0.030~M$_{\sun}$ (i.e.
m$_{C}$~$\geq$~31.5~M$_{Jupiter}$) and a semi-major axis
a$_{C}$~$\geq$~4.4~AU (Table\,4).

\begin{table}
\begin{center}
\caption[]{Orbital solutions for the companion BL~Cam~B.}
\begin{tabular}{l c}  \hline \hline
&   BL~Cam~B\\ \hline LTT  \ semi-amplitude (s)  &   96.6 $\pm$ 0.5
\\
Eccentricity   &   0.19 $\pm$ 0.01\\
Argument \ of \ periastron ($^o$) &   87.5 $\pm$ 2.9 \\
Orbital\  period  (d)&  144.19 $\pm$ 0.02 \\
Periastron \  passage  (BJD)&   2454119 $\pm$  3.831  \\
\hline\
\end{tabular}
\end{center}
\end{table}

\section{Discussion}
\subsection{Frequency of the main pulsation and its variation}

The main frequency pulsation of BL~Cam published in R07 (f$_0$ =
25.57647~d$^{-1}$) was obtained from the Fourier analysis of only a
small part of the time series available for the whole 2005-6
campaign. They considered fewer than 47$\%$ of the total observing
hours, fewer than 31 $\%$ of the total number of nights, and fewer
than 24$\%$ of the total time span of the campaign (128 d over its
total 540 d duration). Their f$_0$ = 25.57647~d$^{-1}$ value (a
period of 0.0390984370 d) was significantly shorter than the one
calculated in this work, this time dealing with all available times
of light maxima. We obtain here f$_0$ = 25.57690~d$^{-1}$ (a period
of 0.0390977846 d, cf. our Eq. 2). This value is identical to that
determined previously (with a smaller amount of data but with a
rather wide time base) in F06. The difference between the two
published values (here and in F06, versus R07) is quite large,
amounting to 0.00043~d$^{-1}$, i.e., 1 cycle per 6.37~yr.

The increasing rate of change in the main pulsation period of BL~Cam
estimated in Sect. 3 is based on a parabolic fit to the O-C diagram.
Even if the existence of period jumps in the O-C diagram cannot be
ruled out $\it{a ~priori}$ (F06), the large and positive rate of
period change found here, i.e., $\frac{dP}{Pdt}$ = (4.41 $\pm$ 0.06)
x 10$^{-10}$ d$^{-1}$ = (161 $\pm$ 3) x 10$^{-9}$ yr$^{-1}$ is much
larger than those predicted by the evolutionary tracks of SX~Phe
pulsators (Rodr\'{i}guez 2003).

\subsection{ Predicted properties of the companions of BL~Cam~A }

The masses of the two companions of BL~Cam~A estimated in Sect. 4
are the minimum values derived from a Keplerian orbit. Of course,
these calculations of the masses depend strongly on the system
inclination.

For a triple system and an eccentricity of 0.7 for BL~Cam~C orbit,
Table\,4 shows that the possible mass range of BL~Cam~B, between
0.46 and 0.99 M$_{\sun}$, corresponds to inclinations of between 90
and 35$\degr$, and semi-major axes between 0.61 and 0.68 AU. Hence
BL~Cam~B could be either a white or a red dwarf, i.e., in the case
of a red dwarf, a M-type star that would be 4-5 mag fainter than
BL~Cam~A in the V-band. Although this rather massive companion is in
the vicinity of BL~Cam~A, a Roche lobe geometry approximation (e.g.,
Warner 1995; Hameury 2007) shows that the two bodies should be a
detached binary system.

In the case of a synchronized orbit -- for which the orbital period
(144.2~d) is equal to the rotational period of the star --, we can
obtain a minimal value of the equatorial rotational velocity of
BL~Cam~A. With a 1.16~R$_{\sun}$ radius (McNamara 1997), we obtain
0.4~km~s$^{-1}$. This estimation is much lower than the uncertain
upper limit $\it{v}\,  sin\, \it{i}$~$\leq$~18~km~s$^{-1}$ (where
$\it{v}$ is equatorial rotation velocity, and $\it{i}$ is
inclination angle of a spin axis to our line of sight) given by
McNamara (1985) from spectroscopic data. Both values yield a low
rotational velocity for BL~Cam~A.

\begin{table}
\begin{center}
\caption[]{Predicted properties of the two companions of BL~Cam~A.}
\begin{tabular}{c c c c c c}  \hline \hline
&  BL~Cam~B&&                & BL~Cam~C&   \\ \hline
 m$_B$            &   i$_B$  &   a$_B$   & m$_C$            &  i$_C$         & a$_C$     \\
 (M$_{\sun}$)     &   ($\degr$)&   (AU)&  (M$_{\sun}$)     & ($\degr$)      & (AU)    \\ \hline
  0.99&35.0&   0.68&   0.99  &  2.7        & 5.5      \\
  0.50&69.8 &  0.62&   0.50 &  4.4      & 5.0  \\
  0.46 &90&  0.61&   0.08 &  22.9& 4.5   \\
&&&   0.03 &  90& 4.4   \\
\hline
\end{tabular}
\end{center}
\end{table}

In the same way, Table\,4 shows that the suspected component
BL~Cam~C could be either a brown dwarf (of mass between 0.03 and
0.08 M$_{\sun}$ for an inclination of between 90 and 23$\degr$), a
low-mass red star (of mass between 0.08 and 0.5 M$_{\sun}$ for an
inclination of between 23 and 4$\degr$), or even a more massive star
i.e., a white dwarf (of mass between 0.5 and 1 M$_{\sun}$ for an
inclination of between 4 and 2.5$\degr$). In these three cases, the
semi-major axis should be, respectively, about 4.5, 4.5 to 5.0, and
5.0 to 5.5~AU. A random distribution of orbital inclinations yields
a higher 0.92 probability of a brown dwarf, and a lower 0.077
probability of a low mass red star, while a more massive companion
seems very unlikely (probability 0.0015).

If BL~Cam~C were a brown dwarf, it should be located in the ``brown
dwarf desert'', given its 4.5 AU separation from BL~Cam~A. This
``brown dwarf desert'' is not yet a well understood region, in which
brown dwarf companions located within a few astronomical units of
their host stars are not detected (e.g., Grether \& Lineweaver 2006;
K\"{u}rster et al. 2008 and references therein).

\subsection{ Hypothesis about the formation of the BL~Cam system}

It is impossible to know whether the system formed as a multiple
(binary or triple) system, or its companions have been driven around
BL~Cam~A by a capture mechanism. In the case of a third component
(BL~Cam~C), its probable low mass, long period, and possible high
eccentricity favours a capture process (e.g., Halbwachs 2007). As
stellar collisions are very uncommon in the halo field (e.g.,
Preston \& Sneden 2000), the system might have formed together by
means of the fragmentation of a collapsing cloud or disk, about
5.4~Gyr ago (McNamara 1997). This scenario would be in agreement
with: (i) the low eccentricity (0.19 $\pm$ 0.01, cf. Table\,3) of
BL~Cam~B, possibly resulting from the tidal circularization of its
orbit (Claret \& Cunha 1997); (ii) the respective closer ($<$ 0.68
AU) and larger ($>$ $\sim$4.5 AU) distances of BL~Cam~B and BL~Cam~C
from the main star. In the case of a triple system, this scenario is
consistent with a stable three-body system configuration over a Gyr
timescale (e.g., Harrington 1968; Szebehely \& Zare 1977;
Quirrenbach 2006).

Assuming a simultaneous formation of the components of the BL~Cam
system, one can argue that the orbits of BL~Cam~B and BL~Cam~C
should be coplanar -- even though the actual observations infer
these preferential values for the inclination  (cf. Tokovinin 2008).
In this case, the possible inclination of BL~Cam~C would be
restricted to the range of inclinations permitted for BL~Cam~B,
i.e., from 35 to 90$\degr$. In this case, the range of possible
BL~Cam~C masses is from 0.030 to 0.053 M$_{\sun}$ (i.e., 32 to 56
M$_{Jupiter}$).

\subsection{Multiplicity among the field SX~Phe stars and eventual connection
of BL~Cam with the blue straggler objects }

Since many SX~Phe-type stars have been discovered among the blue
stragglers of globular clusters, the possible connection between
these two populations has been analysed by several authors (e.g.,
Nemec \& Mateo 1990).

The blue stragglers (Sandage 1953) have been found in the halo of
the Milky Way, in globular and open clusters, and in dwarf galaxies.
Their formation and evolutionary status remains unclear. To explain
the increase in the lifetime of these main sequence stars, several
hypotheses have been proposed: the most probable is that the blue
stragglers are interacting binaries, produced by direct stellar
collisions or binary evolution (see, e.g., reviews by Livio 1993 and
Stryker 1993), but they could also be the product of primordial
hierarchical triple stars (Perets \& Fabrycky 2009; see also the
results of Ferraro et al. 2009 and Mathieu \& Geller 2009).

Over the past two decades, large photometric and radial velocity
surveys of metal-poor blue stars (i.e., halo field blue stragglers,
following Jorissen \& Frankowski 2008) have been carried out, mainly
by Preston \& Sneden (2000, and references therein) and Carney et
al. (2005, and references therein). They have led to the discovery
of three new SX~Phe-type stars. Until now, fourteen field
SX~Phe-type stars have been identified (thirteen in Table\,1 of
Rodr\'{i}guez 2003, and CS 22871-040 mentioned in Preston \& Sneden
2000). Three of these field SX~Phe are binary or multiple stars: KZ
Hya, CS 22966-043, and now BL~Cam. Both CS 29499-057 and CY Aqr are
possible binaries, although this requires confirmation (for more
details, see Liu et al. 1991; McNamara 1997, Preston \& Sneden 2000;
Fu \& Sterken 2003; Rodr\'{i}guez 2003; Fu et al. 2008b). This
reduced sample cannot provide robust statistics in our analysis, but
it raises a question about the connection between the SX~Phe stars
and the blue straggler objects as interacting binaries. The orbital
periods are typically some hundreds of days, and the three multiple
known systems have low eccentricities. According to Preston \&
Sneden (2000), these characteristics are consistent with a mass
transfer model.

One may speculate about the observational difference for field stars
between a Population II star (low metallicity, evolved) and a more
recent binary system where the main component has been cannibalized
by its companion, leading to eventual metallic depletion. On the one
hand, BL~Cam has the low metallicity of all field SX~Phe variables,
and the multiplicity expected of blue stragglers in the more usual
hypotheses, but, being a field star, no conclusion can be reach
about its age and evolution.

\section{Conclusion and future prospects}

From 2005 to 2009, an intensive photometric campaign has been
dedicated to the SX~Phe-type star BL~Cam. This campaign has allowed
a large amount of observations to be gathered by amateurs and
professionals from Europe and North-America. This multisite campaign
has led to a new pulsation analysis of BL~Cam. An analysis of the
O-C diagram has been performed to investigate the main pulsation
period of the star. A short period (144.19 $\pm$ 0.02 d) has been
detected in the O-C variations, as well as a possible long-term
periodic (3390 $\pm$ 26 d) variation. We therefore invoked a light
travel-time effect to interpret the short-term variation, and
proposed the presence of a rather massive stellar companion (from
$\sim$0.5 to 1~M$_{\sun}$), closer than $\sim$0.7 AU to BL~Cam, with
a short period orbit (144.2 d). In addition, about two long cycles
of quasi-periodic fluctuations have been observed over a period of
about two decades. This long timescale O-C modulation could also be
interpreted as a light travel-time effect, which is indicative of an
additional object orbiting BL~Cam. If real, it should be a low mass
object ($\geq$ 0.03 M$_{\sun}$), very probably a brown dwarf,
orbiting in $\sim$9.3~yr within a $\sim$4.5 AU distance.

The analysis of the O-C diagram has also allowed us to measure a
positive secular increase in the main period at the rate (4.41 $\pm$
0.06) x $10^{-10}$ d$^{-1}$ = (161 $\pm$ 3) x $10^{-9}$ yr$^{-1}$. A
continuous photometric monitoring is essential for obtaining a
superior knowledge of the oscillations and behaviour of the main
pulsation period of BL~Cam, and confirming the three-body system
hypothesis.

BL~Cam is quite distant ($\sim$1000 pc; McNamara 1997), and its
companion(s) are too faint and too closely orbiting to be observed
by direct imaging techniques. The spectrum of BL~Cam has only a few
lines -- mostly of hydrogen (McNamara \& Feltz 1978; Alvarez \&
Fox-Machado, private comm). From the above orbital parameters, we
can infer that the radial velocity curve should have a
semi-amplitude of $\sim$15 km s$^{-1}$. This could be within the
reach of modern spectrographs, provided we achieve a high enough
signal-to-noise ratio to unambiguously detect and identify lines.

\begin{acknowledgements} E.G.H. acknowledges the use of the 1.8-m Plaskett
Telescope at the Dominion Astrophysical Observatory, Herzberg
Institute of Astrophysics, National Research Council of Canada.
J.H.S. is very grateful to his group of students at the Physics
Department of the Virginia Polytechnic Institute and State
University. LFM and MA acknowledge financial support from the UNAM
under grant PAPIIT~IN114309. SF thanks Russ Robb (University of
Victoria, Canada) who provided the unpublished times of light
maxima, and J\'{e}r\^{o}me Berthier and Fran\c{c}ois Colas (institut
de m\'{e}canique c\'{e}leste et de calcul des
\'{e}ph\'{e}m\'{e}rides, France). This investigation was partially
supported by the Junta de Andaluc\'\i a and the Spanish Direcci\'on
General de Investigaci\'on (DGI) under projects AYA2006-06375 and
AYA2009-10394. The observations were partially collected with the
1.5m telescope at SNO which is operated by the Instituto de
Astrof\'\i sica de Andaluc\'\i a. Data from Pic du Midi Observatory
have been obtained with the 0.6 m telescope, a facility operated by
observatoire Midi-Pyr\'{e}n\'{e}es and Association T60, an amateur
association. Part of the data used in this work was acquired with
equipment purchased thanks to a research fund financed by the
Belgian National Lottery (1999). We acknowledge with thanks the
variable star observations from the American Association of Variable
Star Observers international database contributed by observers
worldwide and used in this study. This research has made use of both
the Simbad database, operated at CDS, Strasbourg, France, and the
Astrophysics Data System, provided by NASA, USA.
\end{acknowledgements}


\begin{thebibliography}{}


\bibitem[\protect\citeauthoryear{Bastian}{2000}]{Bastian}Bastian, U. 2000, IBVS, 4822
\bibitem[\protect\citeauthoryear{Breger}{2000}]{Breger2000} Breger, M. 2000, ASP Conf. Ser., 210, Delta Scuti and
related stars: reference handbook and proceedings of the 6th Vienna
workshop in astrophysics, ed. M. Breger \& M.H. Montgomery,
Astronomical Society of the Pacific, 3
\bibitem[\protect\citeauthoryear{Breger \& Pamyatnykh}{1998}]{Breger} Breger, M., \& Pamyatnykh, A.A. 1998, A\&A, 332, 958
\bibitem[\protect\citeauthoryear{Carney et al.}{2005}]{Carney} Carney, B.W., Latham, D. W., \& Laird, J.B. 2005, AJ, 129, 466
\bibitem[\protect\citeauthoryear{Chandler}{1892}]{Chandler} Chandler, S.C. 1892, AJ, 11, 113
\bibitem[\protect\citeauthoryear{Claret \& Cunha}{1997}]{Claret}Claret, A., \& Cunha, N.C.S. 1997, A\&A, 318, 187
\bibitem[\protect\citeauthoryear{Debarbat}{1978}]{Debarbat} D\'ebarbat, S. 1978, in R\"{o}emer et la vitesse de la
lumi\`ere, ed. R. Taton, Editions Vrin, 143
\bibitem{Delaney} Delaney, P., Robb, R.M., Berndsen, A., Blake, R.M., \&
Khosravani, H. 2000, IBVS, unpublished
\bibitem[\protect\citeauthoryear{Eddington}{1918}]{Eddington} Eddington, A.S. 1918, MNRAS, 79, 2
\bibitem[\protect\citeauthoryear{Fauvaud et al.}{2006}]{Favaud2006} Fauvaud, S., Rodr\'{i}guez, E., Zhou, A.Y., et al. 2006,
A\&A, 451, 999 (F06)
\bibitem[\protect\citeauthoryear{Ferraro et al.}{2009}]{Ferraro} Ferraro, F.R., Beccari, G., Dalessandro, E., et al. 2009,
Nature, 462, 1028
\bibitem[\protect\citeauthoryear{Fu \& Sterken}{2003}]{Fu2003} Fu, J.N. \& Sterken, C. 2003, A\&A, 405, 685
\bibitem[\protect\citeauthoryear{Fu et al.}{2008a}]{Fu2008a} Fu, J.N., Zhan, C., Marak, K., Boonyarak, C., Khokhuntod,
P., \& Jiang, S.-Y. 2008a, ChJAA, 8, 237
\bibitem[\protect\citeauthoryear{Fu et al.}{2008b}]{Fu2008b} Fu, J.N., Khokhuntod, P., Rodr\'{i}guez, E., et al. 2008b,
AJ, 135, 1958
\bibitem[\protect\citeauthoryear{Grether}{2006}]{Grether} Grether, D., \& Lineweaver, C.H. 2006, ApJ, 640, 1051
\bibitem[\protect\citeauthoryear{Halbwachs}{2007}]{Halbwachs} Halbwachs, J.-L. 2007, Binary stars, in ELSA school on the
science of Gaia,
http://www.astro.lu.se/ELSA/pages/PublicDocuments/Halbwachs.pdf
\bibitem[\protect\citeauthoryear{Hameury}{2007}]{Hameury} Hameury J.-M. 2007, in Limites et lobes de Roche, ed.
J.-M Faidit, Vuibert and Soci\'et\'e astronomique de France, 275
\bibitem[\protect\citeauthoryear{Handler}{2000}]{Handler} Handler, G. 2000, in Variable stars as essential
astrophysics tools, ed. C. Ibano\v{g}lu, NATO Sciences Series C544,
Kluwer Academic Publishers, 557
\bibitem[\protect\citeauthoryear{Harrington}{1968}]{Harrington} Harrington, R.S. 1968, AJ, 73, 190
\bibitem[\protect\citeauthoryear{Hilditch}{2001}]{Hilditch} Hilditch, R.W. 2001, An introduction to close binary stars,
Cambridge University Press
\bibitem[\protect\citeauthoryear{Irwin}{1952}]{Irwin} Irwin, J.B. 1952, ApJ, 116, 211
\bibitem[\protect\citeauthoryear{Jeon et al.}{2003}]{Jeon2003} Jeon, Y.B., Lee, M.G., Kim, S.L., \& Lee, H. 2003, AJ, 125,
3165
\bibitem[\protect\citeauthoryear{Jeon et al.}{2004}]{Jeon2004} Jeon, Y.B., Lee, M.G., Kim, S.L., \&  Lee, H. 2004, AJ,
128, 287
\bibitem[\protect\citeauthoryear{Jorissen}{2008}]{Jorissen} Jorissen, A., \& Frankowski, A. 2008, AIP Conf. Proc.,
1057. Graduate school in astronomy: XII Special courses at the
National Observatory of Rio de Janeiro, ed. P. Pellegrini, S.
Daflon, J. Alcaniz \& E. Telles, American Institute of Physics, 1
\bibitem[\protect\citeauthoryear{Kim et al.}{2003}]{Kim} Kim, C., Jeon, Y.-B., \& Kim, S.-L. 2003, PASP, 115, 755
\bibitem[\protect\citeauthoryear{Klingenberg et al.}{2006}]{Klingenberg} Klingenberg, G., Dvorak, S.W., \& Roberston, C.W. 2006,
IBVS, 5701
\bibitem[\protect\citeauthoryear{K\"{u}rster}{2008}]{Kurster} K\"{u}rster, M., Endl, M., \& Reffert, S. 2008, A\&A, 483, 869
\bibitem[\protect\citeauthoryear{Lenz \& Breger}{2005}]{Lenz} Lenz, P., \& Breger, M. 2005, Comm. Asteroseis., 146, 5
\bibitem[\protect\citeauthoryear{Liu et al.}{1991}]{Liu} Liu, Y.Y., Jiang, S.Y., \& Cao, M. 1991, IBVS, 3606
\bibitem[\protect\citeauthoryear{Livio}{1993}]{Livio} Livio, M. 1993, ASP Conf. Ser., 53, Blue stragglers, ed. R.A.
Saffer, Astronomical Society of the Pacific, 3
\bibitem[\protect\citeauthoryear{Martin}{1996}]{Martin1996} Martin, B. 1996, CCD Astron., Summer 1996, 7
\bibitem[\protect\citeauthoryear{Martin}{2001}]{Martin2001} Martin, B. 2001, http://www.kingsu.ab.ca/$\sim$brian/astro/cba$\_$alta/data$\_$ archive/blcam/blcam.html
\bibitem[\protect\citeauthoryear{Mathieu \&  Geller}{2009}]{Mathieu} Mathieu, R.D., \& Geller, A.M. 2009, Nature, 462, 1032
\bibitem[\protect\citeauthoryear{McCarthy \& Seidelmann}{2009}]{McCarthy2009} McCarthy, D.D., \& Seidelmann, P.K. 2009, Time-From Earth rotation to atomic physics, Wiley VCH
\bibitem[\protect\citeauthoryear{McNamara}{1985}]{McNamara1985} McNamara, D.H. 1985, PASP, 97, 715
\bibitem[\protect\citeauthoryear{McNamara}{1997}]{McNamara1997} McNamara, D.H. 1997, PASP, 109, 1221
\bibitem[\protect\citeauthoryear{McNamara \& Feltz}{1978}]{McNamara1978} McNamara, D.H., \& Feltz, K.A. 1978, PASP, 90, 275
\bibitem[\protect\citeauthoryear{Mazur}{2003}]{Mazur} Mazur, B., Krzeminski, W., \& Thompson, I.B. 2003, MNRAS 340,
1205
\bibitem[\protect\citeauthoryear{Nemec \& Mateo}{1990}]{Nemec} Nemec, J., \& Mateo, M. 1990, ASP Conf. Ser., 11,
Confrontation between stellar pulsation and evolution, ed. C.
Cacciari \& G. Clementini, Astronomical Society of the Pacific, 64
\bibitem[\protect\citeauthoryear{Percy}{2007}]{Percy} Percy, J.R., Bandara, K., \& Cimino, P. 2007, JAAVSO,
http://www.aavso.org/publications/ejaavso
\bibitem[\protect\citeauthoryear{Perets}{2009}]{Perets} Perets, H.B., \& Fabrycky, D.C. 2009, ApJ 697, 1048
\bibitem[\protect\citeauthoryear{Preston \& Sneden}{2000}]{Preston} Preston, G.W., \& Sneden, C. 2000, AJ, 120, 1014
\bibitem[\protect\citeauthoryear{Quirrenbach}{2006}]{Quirrenbach} Quirrenbach, A. 2006, in Extrasolar planets, ed.
M. Mayor, D. Queloz \& S. Udry, Saas-Fee Adv. Course 31, Springer, 1
\bibitem[\protect\citeauthoryear{Ribas}{2005}]{Ribas2005} Ribas, I. 2005, ASP Conf. Ser., 335, The light-time effect
in astrophysics: causes and cures of the O-C diagram, ed. C.
Sterken, Astronomical Society of the Pacific, 55
\bibitem[\protect\citeauthoryear{Ribas et al.}{2002}]{Ribas2002} Ribas, I., Arenou, F., \& Guinan, E.F. 2002, AJ, 123, 2033
\bibitem[\protect\citeauthoryear{Rodr\'{i}guez}{2003}]{Rodriguez2003} Rodr\'{i}guez, E. 2003, Recent Res. Devel. Astron.
Astrophys., 1, 881
\bibitem[\protect\citeauthoryear{Rodr\'{i}guez \&  L\'opez-Gonz\'alez}{2000}]{Rodriguez2000} Rodr\'{i}guez, E., \& L\'opez-Gonz\'alez, M.J. 2000,
A\&A, 359, 597
\bibitem[\protect\citeauthoryear{Rodr\'{i}guez et al.}{2007}]{Rodriguez2007} Rodr\'{i}guez, E., Fauvaud, S., Farrell, J.A. et al.
2007, A\&A, 471, 255 (R07)
\bibitem[\protect\citeauthoryear{Sandage}{1953}]{Sandage} Sandage, A.R. 1953, AJ, 58, 61
\bibitem[\protect\citeauthoryear{Seidelmann \& Fukushima}{1992}]{Seidelmann} Seidelmann, P.K., \& Fukushima, T. 1992, A\&A, 265, 833
\bibitem[\protect\citeauthoryear{Souchay}{2007}]{Souchay} Souchay, J. 2007, in Topics in gravitational dynamics:
solar, extra-solar and galactic systems, ed. D. Benest, C.
Froeschl\'e \& E. Lega, Lect. Notes Phys. 729, Springer, 151
\bibitem[\protect\citeauthoryear{Sterken}{2005a}]{Sterken2005a} Sterken, C. 2005a, ASP Conf. Ser., 335, The light-time
effect in astrophysics: causes and cures of the O-C diagram, ed. C.
Sterken, Astronomical Society of the Pacific, 3
\bibitem[\protect\citeauthoryear{Sterken}{2005b}]{Sterken2005b} Sterken, C. 2005b, ASP Conf. Ser., 335, The light-time
effect in astrophysics: causes and cures of the O-C diagram, ed. C.
Sterken, Astronomical Society of the Pacific, 181
\bibitem[\protect\citeauthoryear{Stryker}{1993}]{Stryker} Stryker, L.L. 1993. PASP, 105, 1081
\bibitem[\protect\citeauthoryear{Szebehely \& Zare}{1977}]{Szebehely} Szebehely, V., \& Zare, K. 1977, A\&A, 58, 145
\bibitem[\protect\citeauthoryear{Szeidl}{2005}]{Szeidl} Szeidl, B. 2005, ASP Conf. Ser., 333, Tidal evolution and
oscillations in binary stars: third Granada workshop on stellar
structure, ed. A. Claret, A. Gim\'enez \& J.-P. Zahn, Astronomical
Society of the Pacific, 183
\bibitem[\protect\citeauthoryear{Templeton}{2005}]{Templeton} Templeton, M.R. 2005, JAAVSO, 34, 1
\bibitem[\protect\citeauthoryear{Tokovinin}{2008}]{Tokovinin} Tokovinin, A. 2008, MNRAS 389, 925
\bibitem[\protect\citeauthoryear{Vanmunster}{2008}]{Vanmunster} Vanmunster, T. 2008, Peranso-Light curve and period
analysis software, http://www.peranso.com/
\bibitem[\protect\citeauthoryear{Warner}{1995}]{Warner} Warner, B. 1995, Cataclysmic variable stars, Cambridge
University Press
\bibitem[\protect\citeauthoryear{Zhou et al.}{1999}]{Zhou} Zhou, A.-Y., Rodr\'{i}guez, E., Jiang, S.-Y., Rolland, A., \&
Costa, V. 1999, MNRAS, 308, 631

\end{thebibliography}

\end{document}